\def \SAIT #1 #2 {{\em Mem.\ Soc.\ Astron.\ It.\/} {\bf #1}, #2}
\def \MESS #1 #2 {{\em The Messenger\/} {\bf #1}, #2}
\def \ASTRNACH #1 #2 {{\em Astron. Nach.\/} {\bf #1}, #2}
\def \AAP #1 #2 {{\em Astron. Astrophys.\/} {\bf #1}, #2}
\def \AAL #1 #2 {{\em Astron. Astrophys. Lett.\/} {\bf #1}, L#2}
\def \AAR #1 #2 {{\em Astron. Astrophys. Rev.\/} {\bf #1}, #2}
\def \AAS #1 #2 {{\em Astron. Astrophys. Suppl. Ser.\/} {\bf #1}, #2}
\def \AJ #1 #2 {{\em Astron. J.\/} {\bf #1}, #2}
\def \ANNREV #1 #2 {{\em Ann. Rev. Astron. Astrophys.\/} {\bf #1}, #2}
\def \APJ #1 #2 {{\em Astrophys. J.\/} {\bf #1}, #2}
\def \APJL #1 #2 {{\em Astrophys. J. Lett.\/} {\bf #1}, L#2}
\def \APJS #1 #2 {{\em Astrophys. J. Suppl.\/} {\bf #1}, #2}
\def \APSS #1 #2 {{\em Astrophys. Space Sci.\/} {\bf #1}, #2}
\def \ASR #1 #2 {{\em Adv. Space Res.\/} {\bf #1}, #2}
\def \BAIC #1 #2 {{\em Bull. Astron. Inst. Czechosl.\/} {\bf #1}, #2}
\def \JSQRT #1 #2 {{\em J. Quant. Spectrosc. Radiat. Transfer\/} {\bf #1}, #2}
\def \MN #1 #2 {{\em Mon. Not. R. Astr. Soc.\/} {\bf #1}, #2}
\def \MEM #1 #2 {{\em Mem. R. Astr. Soc.\/} {\bf #1}, #2}
\def \PLR #1 #2 {{\em Phys. Lett. Rev.\/} {\bf #1}, #2}
\def \PASJ #1 #2 {{\em Publ. Astron. Soc. Japan\/} {\bf #1}, #2}
\def \PASP #1 #2 {{\em Publ. Astr. Soc. Pacific\/} {\bf #1}, #2}
\def \NAT #1 #2 {{\em Nature\/} {\bf #1}, #2}
\def\farcs{\hbox{$.\!\!^{\prime\prime}$}}
\def\fdg{\hbox{$.\!\!^\circ$}}
\def\kms{km~s$^{-1}$}
\def\la{\ifmmode\stackrel{<}{_{\sim}}\else$\stackrel{<}{_{\sim}}$\fi} 
\def\ga{\ifmmode\stackrel{>}{_{\sim}}\else$\stackrel{>}{_{\sim}}$\fi} 

\documentstyle[epsfig]{memsait}
\input epsf.sty
\begin{opening}
\title{A SEARCH FOR PULSAR WIND NEBULAE USING PULSAR GATING} 
\author{B. M. GAENSLER$^{1,2}$\thanks{Current address:
Center for Space Research, Massachusetts Institute of Technology,
MA, USA. Email: bmg@space.mit.edu}, 
B. W. STAPPERS$^3$, D. A. FRAIL$^4$, S. JOHNSTON$^5$}
\institute{$^1$Astrophysics Department, School of Physics, University
of Sydney, NSW, Australia \\
$^2$Australia Telescope National Facility, CSIRO, NSW, Australia \\
$^3$Astronomical Institute ``Anton Pannekoek'', University of
Amsterdam, The Netherlands \\
$^4$National Radio Astronomy Observatory, NM, USA \\
$^5$Research Centre for Theoretical Astrophysics, University of Sydney,
NSW, Australia}
\date{} 
\end{opening}

\begin{document}

\oddpagefooter{}{}{} 
\evenpagefooter{}{}{} 
\ 
\bigskip

\begin{abstract}
Previous searches for radio pulsar wind nebulae (PWN) suggest that only
the youngest and most energetic pulsars power PWN.  However the
selection effects associated with such searches are severe, in that
emission from a faint compact PWN will be masked by that from its
associated pulsar. This has motivated us to search for radio PWN using
the technique of pulsar gating, which allows us to ``switch off'' the
pulsed emission and see what is hidden underneath.  This search has
resulted in the detection of the faintest radio PWN yet discovered,
while non-detections around other pulsars have implications for
previous claims of nebular emission and for associations with nearby
supernova remnants.
\end{abstract}

\section{Introduction}

A pulsar loses most of its rotational energy in a relativistic wind
which, when confined, may be observable as a pulsar wind nebula (PWN).
PWN are our main diagnostics of the pulsar wind and its environment,
but at radio wavelengths, just seven pulsars are known to have definite
PWN. All these pulsars
are young (six are younger than 20\,000~kyr), have high spin-down
luminosities ($\dot{E} > 10^{35}$~erg~s$^{-1}$) and are associated
with supernova remnants (SNRs). The obvious question to ask is whether there
are radio PWN around other pulsars. However, finding such PWN presents an
observational challenge, because the expected faint and/or compact sources 
might be masked by emission from their associated pulsars.

The results of a systematic survey for radio PWN have been 
reported recently by Frail \& Scharringhausen (1997, hereafter FS97). They used
the Very Large Array (VLA) at 8.3~GHz and at 0\farcs8 resolution,
reasoning that this choice of observing parameters would allow them to
best separate out any PWN from its associated pulsar.  FS97 observed 35
pulsars, quite sensibly choosing pulsars with a high $\dot{E}$ or space
velocity. However, they failed to detect any PWN down to a 5$\sigma$
sensitivity of 0.2~mJy~beam$^{-1}$.  FS97 concluded that only young
energetic pulsars produce observable radio PWN, but could not rule out
that they had simply resolved out any nebular emission.

\section{New Observations}

\begin{figure}
\centerline{\epsfig{file=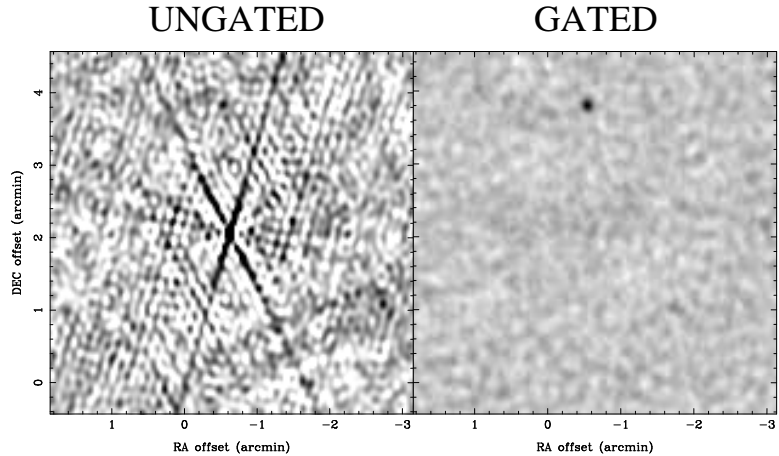,height=6cm}}
\caption[h]{Ungated {\em (left)}\ and gated {\em (right)}\ 1.3~GHz
observations of PSR~B1055--52, both shown with the same grey-scale
range. Without pulsar-gating, the pulsar
itself, as well as artifacts associated with scintillation effects,
make it impossible to say anything useful about the presence or absence
of any PWN. With the pulsed emission removed, we can put a 
limit of 0.4~mJy~beam$^{-1}$ (5$\sigma$) on any unpulsed emission.}
\label{fig_1055_on_off}
\end{figure}

The results of FS97 have prompted us to search for PWN at southern
declinations using the Australia Telescope Compact Array (ATCA). Rather
than just image the pulsar and surrounds, we have used the technique of
{\em pulsar gating}, in which data are recorded at high time resolution
so that images can be made in which the pulsar has been ``switched
off'' (see Figure~\ref{fig_1055_on_off}). We observed five pulsars, at
significantly lower frequency (1.3~GHz) and resolution ($20''$) than
did FS97. Because we can gate out the pulsar, our choice of observing
parameters results in a surface brightness sensitivity {\em 200 times
fainter}\ than that achieved by FS97.

Of the five pulsars we observed with the ATCA, we were successful in
finding a PWN associated with PSR~B0906--49, but saw no nebular
emission towards PSRs B1046--58, B1055--52, J1105--6107 or B1610--50.
Results on PSR~B0906--49 have been reported by Gaensler et al (1998),
while the other pulsars will be discussed fully by Stappers et al
(1998).  Here we review some of the highlights of this study.

\subsection{PSR~B1610--50}

With a characteristic age of just 7.4~kyr, and a high $\dot{E}$ of $2
\times 10^{36}$~erg~s$^{-1}$, PSR~B1610--50 is a prime candidate for
powering a radio PWN.  However a gated image shows
no PWN down to a 5$\sigma$
limit of 1~mJy~beam$^{-1}$.

PSR~B1610--50 is similar in its age, magnetic field strength and
$\dot{E}$ to PSRs~B1757--24 and B1853+01, both of which generate
prominent PWN with bow-shock morphologies.  A bow-shock PWN is confined
by the ram pressure, $\rho v^2$, resulting from the relative motion of
the pulsar with respect to the ISM. The lack of a PWN associated with
B1610--50 suggests that $\rho v^2$ in this case is much less than for
PSRs~B1757--24 and B1853+01, and we thus infer for B1610--50 a
space-velocity $v_{\rm PSR} < 130\, n^{-1/2}$~\kms, where $n$~cm$^{-3}$
is the ambient density.

\begin{figure}
\vspace{-1.5cm}
\centerline{\epsfig{file=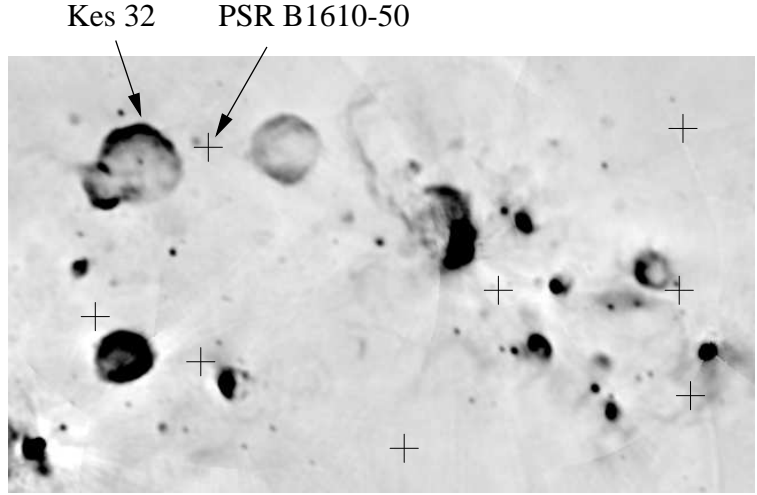,height=10cm}}
\vspace{-2cm}
\caption[h]{843~MHz image of the region surrounding PSR~B1610-50
(Green et al 1998). The x-axis spans 2$^\circ$ of
Galactic longitude, while the y-axis covers $1\fdg2$ of Galactic
latitude. Positions of known pulsars are marked with ``+'' symbols.}
\label{fig_1610_field}
\end{figure}

An association between PSR~B1610--50 and the nearby SNR
Kes~32 has been claimed by Caraveo (1993), but implies a
projected velocity for the pulsar $>$3600~\kms. This can only be
reconciled with our result if $n < 0.001$~cm$^{-3}$, in which case
Kes~32 itself should not be visible. Furthermore, as demonstrated in
Figure~\ref{fig_1610_field} the region is a complex one, containing
three spiral arms, at least three SNRs and eight known pulsars. Thus
spatial proximity alone is hardly compelling evidence for a pulsar/SNR
association.  Finally, if PSR~B1610--50 indeed has this high inferred space
velocity, then one might expect significant distortion of Kes~32
where it was penetrated by the pulsar (as is seen in the case
of  G5.4--1.2 and
PSR~B1757--24). However the appropriate part of
the shell shows no such effect. From all these results, we can
conclude that PSR~B1610--50 is {\em not}\ associated with SNR~Kes~32,
nor with any other known remnant. The lack of an associated SNR for such
a young pulsar implies an ambient density $n \la 0.05$~cm$^{-3}$ (e.g.\
Kafatos et al 1980),
which is consistent with the lack of a radio PWN for typical pulsar
space velocities.

\subsection{PSR~B1055--52}

PSR~B1055--52 is a comparatively old pulsar ($\tau = 345$~kyr) with
a moderate spin-down luminosity ($\dot{E} = 3 \times
10^{34}$~erg~s$^{-1}$). An extended radio PWN was reported by
Combi et al (1997) using observations with a 30~m single dish, while
in hard X-rays an extended clumpy PWN is seen by {\em ASCA}\ (Shibata
et al 1997). 

\begin{figure}
\centerline{\epsfig{file=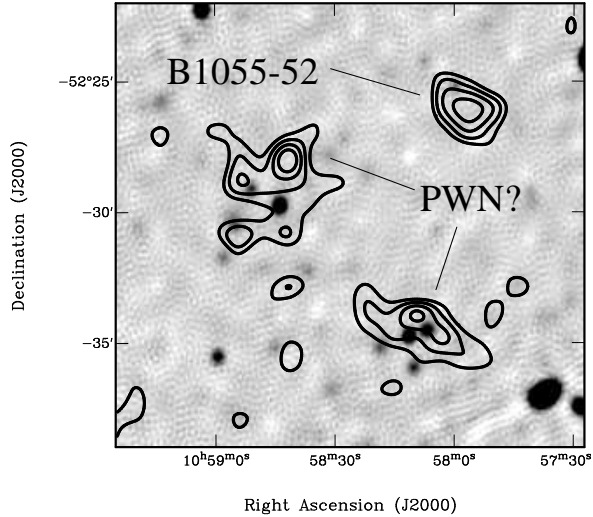,height=8cm}}
\caption[h]{Radio / X-ray comparison of the region surrounding
PSR~B1055--52. The grey-scale represents 20~cm ATCA data, with
the pulsar gated out. The contours correspond to 0.5--2~keV {\em ASCA}\ GIS
data of Shibata et al (1997).}
\label{fig_1055_asca_atca}
\end{figure}

Our gated ATCA data show no radio PWN associated with PSR~B1055--52 on
any spatial scale (see Figures~\ref{fig_1055_on_off} and
\ref{fig_1055_asca_atca}).  Our observations certainly have the
sensitivity and $u-v$ coverage to detect the radio nebula of Combi et
al (1997), and we conclude that the source they claim as a PWN is
clearly spurious and is a result of confusion, typical when small
single dishes are pointed at complex regions.

In Figure~\ref{fig_1055_asca_atca} we show an overlay between radio and
X-ray images of the region.  The clumps of X-ray emission, claimed by
Shibata et al (1997) to be part of an extended PWN, coincide with radio
point sources, and also correspond to X-ray point sources in {\em
ROSAT}\ data (W. Becker, these proceedings).  Thus we think it more likely
that the apparent X-ray PWN corresponds to emission from unrelated
background sources.

\subsection{PSR~B0906--49}

\begin{figure}
\centerline{\epsfig{file=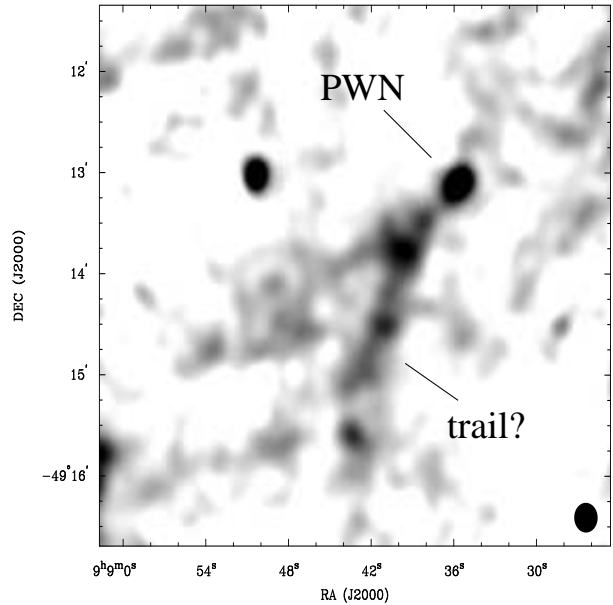,height=8cm}}
\caption[h]{Gated 20~cm image of PSR~B0906--49. The pulsar has been
gated out, but is coincident with the PWN. The FWHM of the synthesised
beam is shown at lower-right.}
\label{fig_0906}
\end{figure}

PSR~B0906--49 is a relatively young ($\tau = 112$~kyr) and energetic
($\dot{E} = 5 \times 10^{35}$~erg~s$^{-1}$) pulsar. Gated ATCA
observations result in the image shown in Figure~\ref{fig_0906}, in which 
an off-pulse source can be seen at the position of the pulsar.
This source is extended and has no detectable polarisation, and we
conclude that it is best interpreted as a bow-shock PWN associated with
PSR~B0906--49. A possible trail extends behind the PWN, with an axis
of symmetry which aligns with the direction along which
the PWN is extended.

Using a projected space velocity for B0906--49 of $\sim$60~\kms\ (Johnston
et al 1998),
pressure balance arguments can be used to infer an ambient density
$n > 2$~cm$^{-3}$. Thus the ram pressure which confines this PWN
results mainly from the high ambient density, rather than from 
the pulsar velocity. PSR~B0906--49 is the oldest pulsar yet to be
associated with a radio PWN; it is also worth noting that this PWN is 100 times
fainter than could have been detected by FS97.

\section{Conclusions}

Pulsar-gating with the ATCA has allowed us to carry out a deep search
for radio PWN around five pulsars. No PWN was found for
PSRs B1046--58, B1055--52, B1105--6107 or B1610--50.
From these results, we argue that PSR~B1610--50 is
not associated with SNR~Kes~32, and also conclude
that there is no evidence to support
previous claims of radio and X-ray nebulae associated with
PSR~B1055--52.
We successfully detected a faint PWN around PSR~B0906--49. This PWN
seems quite different from other bow-shock PWN, in that it
is associated with a low velocity pulsar in a dense medium.

Our results bear out the expectation that only pulsars with a high
$\dot{E}$ will produce radio PWN. Data on known PWN suggest that a
minimum ram pressure $n \times v_{\rm PSR}^2 \ga 10^4$~(\kms)$^2$~cm$^{-3}$
is required to confine the wind sufficiently to produce observable radio
emission.  

Our small survey clearly demonstrates that there are severe selection
effects associated with previous searches for radio PWN.  There may
well be an undiscovered population of faint PWN similar to that found
around PSR~B0906--49, and we intend to follow up on this possibility by
using pulsar-gating on the VLA to observe a large number of sources.
Apart from finding new and interesting PWN, we hope to use the results
of such a survey to properly constrain the minimum $\dot{E}$, $n$ and
$v_{\rm PSR}$ required to produce a radio PWN.

\acknowledgements

BMG is grateful to the Local Organising Committee, the Science
Foundation for Physics within the University of Sydney, the R.\ and
M.\ Bentwich Scholarship and the James Kentley Memorial Scholarship for
financial assistance in attending this workshop. The Australia
Telescope is funded by the Commonwealth of Australia for operation as a
National Facility managed by CSIRO.


\end{document}